\documentclass[aps,floatfix,preprint,nofootinbib]{revtex4}
\usepackage{graphics}
\usepackage{graphicx}
\usepackage{amssymb}		
\usepackage{amsmath}
\usepackage{verbatim}
\usepackage{pstricks}
\usepackage{slashed}

\newcommand{\nc}{\newcommand}
\nc{\beq}{\begin{equation}}
\nc{\eeq}{\end{equation}}
\nc{\barray}{\begin{eqnarray}}
\nc{\earray}{\end{eqnarray}}
\nc{\barrayn}{\begin{eqnarray*}}
\nc{\earrayn}{\end{eqnarray*}}
\nc{\bcenter}{\begin{center}}
\nc{\ecenter}{\end{center}}
\nc{\ket}[1]{| #1 \rangle}
\nc{\bra}[1]{\langle #1 |}
\nc{\0}{\ket{0}}
\nc{\mc}{\mathcal}
\nc{\er}[1]{(\ref{eq:#1})}
\nc{\onehalf}{\frac{1}{2}}
\nc{\partialbar}{\bar{\partial}}
\nc{\psit}{\widetilde{\psi}}
\nc{\Tr}{\mbox{Tr}}
\nc{\ev}{\;\mathrm{eV}}
\nc{\mev}{\;\mathrm{MeV}}
\nc{\gev}{\;\mathrm{GeV}}
\nc{\tev}{\;\mathrm{TeV}}
\nc{\infinity}{\infty}

\def\squark{\widetilde{q}}

\def\chii0{\chi_i^0}
\def\chij0{\chi_j^0}

\newcommand{\gsim}{\lower.7ex\hbox{$\;\stackrel{\textstyle>}{\sim}\;$}}
\newcommand{\lsim}{\lower.7ex\hbox{$\;\stackrel{\textstyle<}{\sim}\;$}}

\begin{document}

\setlength{\baselineskip}{0.22in}

\begin{flushright}MCTP-10-28 \\
\end{flushright}
\vspace{0.2cm}

\title{Darkogenesis}

\author{
Jessie Shelton$^a$\footnote{\tt j.shelton@yale.edu},
Kathryn M. Zurek$^{b}$
}

\vspace*{0.4cm}

\affiliation{
$^a$Department of Physics, Yale University, New Haven, CT 06520 
\\
$^b$Michigan Center for Theoretical Physics, Department of Physics, University of Michigan, Ann Arbor, MI 48109
}

\date{\today}

\begin{abstract}
\noindent

In standard models of baryogenesis and of dark matter, the mechanisms
which generate the densities in both sectors are unrelated to each
other.  In this paper we explore models which generate the baryon
asymmetry through the dark matter sector, simultaneously relating the
baryon asymmetry to the dark matter density.  In the class of models
we explore, a dark matter asymmetry is generated in the hidden sector
through a first order phase transition.  Within the hidden sector, it is easy to achieve a sufficiently strong first order
phase transition and large enough $CP $ violation to generate the observed asymmetry.  This can happen above
or below the electroweak phase transition, but in both cases
significantly before the dark matter becomes non-relativistic.  We
study examples where the Asymmetric Dark Matter density is then
transferred to the baryons both through perturbative and
non-perturbative communication mechanisms, and show that in both cases
cosmological constraints are satisfied while a sufficient baryon asymmetry can be generated.

\end{abstract}

\maketitle

\section{Hidden sector baryogenesis}
\label{sec:intro}

The Standard Model (SM) of particle physics has proven remarkably
successful at describing the phenomena observed at colliders, from the
detailed properties of the gauge sector to flavor physics at bottom
and charm factories.  Despite this unprecedented success, we know that
the SM must be incomplete.  Two fundamental features of the observed
universe cannot be explained within the SM: the presence of the baryon
asymmetry, and the existence of dark matter (DM).  The SM contains
neither sufficient $CP $ violation to produce the observed size of the
baryon asymmetry, nor a particle which can act as DM.

Typically the solutions to these two puzzles are treated
independently.  Observationally it is known that the DM and
baryon densities are approximately the same,
\begin{equation}
\frac{\rho_{DM}}{\rho_b} \approx 5.
\label{density}
\end{equation}
However, in most models the DM and baryon densities are not
directly related to each other.  For example, within the Minimal
Supersymmetric Standard Model (MSSM), the near-equality of dark and
visible relic densities can be accommodated as a coincidence, since
${\cal O}(1)$ $CP$ violating parameters together with a TeV mass
scale can give rise to both a thermal relic abundance of a weakly
interacting DM particle and an asymmetric relic abundance of baryonic
matter, satisfying Eq.~\ref{density} (see {\em e.g.}
\cite{Carena1,Carena2}).  However, within the MSSM, it is just as
natural for the DM and baryon densities separately to be several
orders of magnitude different than their observed values, so that the
MSSM does not explain why dark and visible densities appear to be so
closely related to each other.

An alternate approach is to take the DM relic density to be
{\it asymmetric}, set by the asymmetry between DM and
anti-DM, $n_X - n_{\bar{X}}$, just as the baryonic relic
density.  Relating the DM number asymmetry to baryon number
then provides a solution to the puzzle of why the DM and baryon energy densities are so close to each other.  Models of this
type \cite{ADMmodels,dbk,ADM} (called Asymmetric DM (ADM) by \cite{ADM}) have sharply
different phenomenology than thermal models; in particular,
the natural scale for ADM is several GeV, since
\begin{equation}
m_{DM} = c \frac{\rho_{DM}}{\rho_b} m_p,
\end{equation}
where $c$ is an ${\cal O}(1)$ number whose exact size is set by the
details of the transfer mechanism.  However, much heavier ADM can be
possible if there is a coincidence of scales, such that the DM is
becoming non-relativistic just as the operator relating dark number to
baryon number is decoupling, as in \cite{techni,ADMpam}, or if there is a cancellation between the injected baryon, lepton and dark number asymmetries, as in \cite{quirk}.  We will
construct models which use both of these mass windows.

Tying baryon number to DM number neatly explains the
coincidence problem, but does not in itself address the {\it origin}
of the asymmetry, only its distribution between sectors.  In this
paper we develop models of ADM where the dark
sector itself is responsible for generating the observed
matter-antimatter asymmetry.

In order to generate a nonzero baryon asymmetry, the
Sakharov conditions must be satisfied \cite{Sakharov}:
\begin{itemize}
\item Baryon or lepton number violation
\item Departure from thermal equilibrium
\item $C$ and $CP$ violation.
\end{itemize}  
When the dark sector is responsible for originating the asymmetry,
these conditions become:
\begin{itemize}
\item the hidden sector must furnish a departure from thermal
      equilibrium;
\item  the hidden sector global symmetry stabilizing the DM as well as either baryon ($B $) or lepton ($L $) number
      must be broken by one or more hidden sector processes, while
      the couplings between the dark sector and the SM
      must conserve a linear combination of the dark global symmetry, $B$, and/or $L$;
\item both $C$ and $CP$ must be violated in the dark sector. 
\end{itemize}
We choose to focus on the scenario where the the departure from
equilibrium is provided by a first-order phase transition in the dark
sector, in which the SM does not directly participate.
The observed matter-antimatter asymmetry is then entirely generated at
this phase transition.  We describe the minimal hidden sector which is
capable of meeting the Sakharov criteria for generating the asymmetry 
in the hidden sector.  After the asymmetry is generated, it must be redistributed from the 
dark sector to the visible sector.  The mechanism which transfers the
asymmetry to the SM must necessarily break either baryon or 
lepton number, and can be either perturbative or nonperturbative.  We 
find the most stringent constraints, perhaps not surprisingly, arise 
from the transfer mechanisms. 

Finally, the hidden sector must satisfy some additional constraints in order
to yield a satisfactory cosmology.  In particular, the symmetric portion
of the DM abundance must annihilate efficiently away \cite{gevhidden}, and the
contribution of hidden sector states to the expansion of the universe
must be minimal by the time of nucleosynthesis; both of these conditions 
can be satisfied by including additional, baryon-, lepton- and dark-number preserving 
couplings between the hidden sector and the SM.

\begin{figure}[t]
\begin{center}
\includegraphics[width=0.8\textwidth]{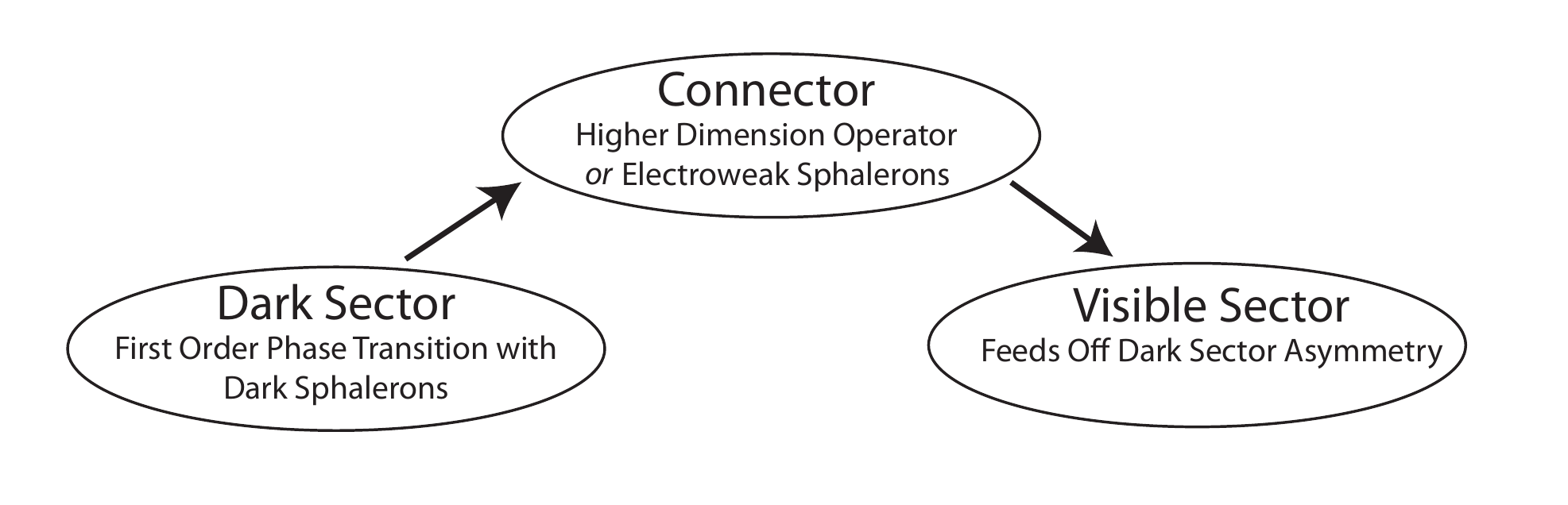}
\end{center}
\caption{A schematic of the classes of models we consider.  The asymmetry is generated in 
the hidden sector via a first order phase transition and then transferred to the visible 
sector either via a higher dimension operator or electroweak sphalerons.}
\label{fig1}
\end{figure}

We show in Fig.~(\ref{fig1}) a schematic of the classes of models we
will consider in this paper.  We begin in section~\ref{sec:HS} by building 
the minimal hidden
sector which accomplishes darkogenesis.  As the dark sector is
relatively unconstrained, most of the constraints reside in the
mechanism chosen for transferring the asymmetry between sectors, which we
discuss in section~\ref{sec:transfer}.  In section~\ref{sec:models} we
build two explicit
models based on the minimal hidden sector, using both perturbative and
nonperturbative mechanisms for transferring the asymmetry between
sectors.  In section~\ref{sec:conclusion}, we conclude.

\section{A First-Order Dark Phase Transition}
\label{sec:HS}

Models where the matter-antimatter asymmetry are generated from a
first-order phase transition are an elegant approach to generating the
observed matter abundance.  The SM famously includes in
principle all of the ingredients necessary for a baryon asymmetry to
be generated at the electroweak phase transition, but quantitatively
fails to generate the observed baryon excess.  Much work has been
performed on extending the matter content of the SM to increase both
the magnitude of the $CP $ violation and the departure from
equilibrium in order to rescue electroweak baryogenesis.  Some of these models also
include a DM candidate, whether thermal, as in the (n)MSSM \cite{Menon},
or non-thermal, as in \cite{Das:2009ue,Craig:2010au}.

We choose to focus on the less-studied scenario where the dark sector
itself has a first-order phase transition in which the SM
does not directly participate.  In this scenario, the necessary
departure from equilibrium occurs due to the passage of supercritical
bubbles of the broken phase through the plasma.  In order to generate
a matter-antimatter asymmetry, the other two Sakharov conditions must
additionally be satisfied.  These can be satisfied naturally within chiral
non-Abelian hidden sectors.

First, there must be a dark number-violating process which is
efficient in the symmetric phase and suppressed in the broken phase,
and which moreover shuts off faster than the timescale for the passage
of the bubble wall.  The requirement for the dark number-violating
process to shut off precisely at the phase transition is nontrivial.
The most natural candidate process is dark sphalerons, the rate for which becomes
exponentially suppressed precisely at the phase transition.  Thus
we achieve dark baryogenesis via the symmetry-breaking phase
transition of a dark non-Abelian gauge group, $SU(\chi)$, and require
that the dark sector contain matter fields which have a global
symmetry $U (1)_D $ which is anomalous under $SU(\chi)$.

To summarize our scenario, we have the following:
\begin{itemize}
\item If the dark sector matter fields are chiral under the dark gauge 
group $SU(\chi)$, this can give rise to an anomalous dark number symmetry $U
(1)_D $ under the $SU(\chi)$.  Dark number violation is then achieved by
dark sphalerons.
\item The group $SU(\chi)$ undergoes a symmetry-breaking first-order
phase transition, during which time a dark matter number asymmetry is generated through the
$C$- and $CP $-violating interactions of the dark sector matter fields with the bubble walls.
\item The asymmetry generated during the phase transition is
transferred to the SM through one of the mechanisms
outlined in section~\ref{sec:transfer}.
\end{itemize}
In our scenario $C$ violation in the dark sector is implemented
through the requirement that the global dark number $U (1)_D $ has a
chiral anomaly under the dark gauge group.

We will take the hidden sector to be weakly coupled for simplicity.
Then the breaking of $SU(\chi)$ is accomplished through a fundamental
scalar Higgs boson, and the interactions of the dark sector fields with the
bubble wall take the form of chiral, $CP$ violating couplings to the
dark Higgs.  The minimal field content in the dark sector which can
satisfy all conditions consists of $2m$ fermionic doublets of an
$SU(2)$ gauge group, which have an anomalous number symmetry $U(1)_D$;
$2\times 2m$ fermionic singlets to form Dirac fermions after
$SU(2)$-breaking; and $k$ Higgs doublet scalars.  The number of
fermionic $SU(2) $ doublets must be even to cancel the global
anomaly. In order to yield a physical $CP$ violating phase, we must
have either $m>1$ or $k>1$, so the minimal matter content consists of
a two Higgs doublet model together with 2 fermionic doublets and 4
fermionic singlets.  Majorana mass terms $M_2 L_DL_D$ for doublets and
$M_1 \bar X \bar X$ for singlets are forbidden by the global dark number
$U(1)_D$.  This minimal hidden sector is summarized in
Table~\ref{table:SU(2)content}.

Since we are utilizing a fundamental scalar, it is appealing to embed
these models within supersymmetry so as to make this hidden sector (as
well as the visible sector) technically natural.  The simplest dark supersymmetric
sector which is chiral under $SU(2)_D$, non-anomalous, and gives
masses to all states in the dark sector is obtained simply by
supersymmetrizing the minimal hidden sector of
Table~\ref{table:SU(2)content}, and is described by the superpotential
\begin{equation}
W = \mu_D H H^c + y_{1i} L_D H \bar{X}_{i} + y_{2j} L_D H^c \bar{X}_{j},
\label{darksuper}
\end{equation}
where $i,~j=1,~2$; we suppress generational indices in
Eq.~(\ref{darksuper}).  

\begin{table}
\begin{center}
\begin{tabular}{ccc}
\hline \hline
                       & $SU(2)_D$ &  $U (1)_D$  \\
\hline
 $H,~H^c$         & 2       &  0\\

 $L_D \times 2$        & 2       &  1\\
 $ \bar{X}_{1,2} \times 2$   & 1       & -1\\
\hline \hline
\end{tabular}
\caption{Matter content in the minimal hidden sector which generates a
matter-antimatter asymmetry from a first-order phase transition.
\label{table:SU(2)content} }
\end{center}
\end{table}

This minimal dark sector is the simplest (supersymmetric) hidden
sector which realizes departure from equilibrium via a first-order
phase transition and satisfies all of Sakharov's criteria.  It will
serve as the basis for our models in section~\ref{sec:models}.  It is
straightforward to ensure that the phase transition is sufficiently first-order
given the unconstrained dynamics in the hidden sector, as we will see
below.  The observable phenomenology of models built on this hidden sector
is dominated by the communication of
the dark baryon number to the visible sector, which we turn to next.

\section{Asymmetry Communication Mechanisms}
\label{sec:transfer}

Once an asymmetry is generated in the hidden sector, it must be
transferred to the SM.  This may be accomplished either
perturbatively or nonperturbatively.  We enumerate the possibilities
and comment on these transfer mechanisms, as well as on the constraints
which each transfer mechanism imposes on the phase transition in the hidden
sector.

\subsection{Higher Dimension Operators}

In order to successfully transfer a dark number asymmetry to a baryon
number asymmetry, the dark sector and the visible sector must be
linked by some mediator states which carry both hidden sector and
SM (gauge and/or global) charges.  There are many
possibilities for these new degrees of freedom.  Below the mass scale
of these new mediators, however, the dark-visible interactions which
they induce can be described using higher-dimension operators in the
effective theory below the mediator mass scale $\Lambda $.  These
contact interactions will generically remain in equilibrium for a
range of temperatures below the mediator mass scale,
allowing us to describe a broad range of possible dark-visible
interactions in terms of a small set of higher-dimensional operators
which carry non-zero $B$ or $L$ as well as non-zero dark number $D $.
The operator in the SM sector must also be sterile, since the DM
itself is sterile.  The lowest-order such operators are
\begin{eqnarray}
{\cal O}_{d+5/2} & = & \frac{{\cal O}_d LH}{\Lambda^{d-3/2}}, \\
{\cal O}_{d+9/2, B} & = & \frac{{\cal O}_d u^cd^cd^c}{\Lambda^{d+1/2}}, \nonumber \\ 
{\cal O}_{d+9/2, L1} & = &\frac{{\cal O}_d L L e^c}{\Lambda^{d+1/2}}, \nonumber \\ 
{\cal O}_{d+9/2, L2} & = &\frac{{\cal O}_d L Q d^c}{\Lambda^{d+1/2}}, \nonumber \\
{\cal O}_{d+5} & = & \frac{{\cal O}_d LH LH}{\Lambda^{d+1}}, \nonumber 
\label{higherdimops}
\end{eqnarray}
where $d$ is the dimension of the dark sector operator, ${\cal O}_d$.
The operator $ {\cal O}_{d+5/2}$ is a special case, and if ${\cal O}_d$
corresponds to a single (fermionic) state in the dark sector, it can
contribute directly to the neutrino mass matrix after electroweak
symmetry breaking; such contributions are highly constrained.

In order to successfully transmit a matter-antimatter asymmetry from a
hidden sector to the SM using one of the operators of
Eq.~\ref{higherdimops}, the hidden sector phase transition must happen
above the temperature $T_f $ at which the operator freezes out.
Moreover, if the hidden sector couples to the visible sector only
through lepton number-violating operators, then the hidden sector
phase transition must occur above the electroweak phase transition, so
that an initial lepton asymmetry can be reprocessed into a baryon
asymmetry.  If, however, the coupling between the sectors proceeds
through the baryon-number violating operator ${\cal O}_{d+9/2, B}$, then the phase transition can occur at
lower scales.  We will construct a low-scale darkogenesis model using
this operator in section~\ref{sec:low} below.

In a supersymmetric theory, the operators of Eq.~\ref{higherdimops}
can be understood as contributions to the superpotential,
\begin{eqnarray}
{\cal O}_{d+2} & = & \frac{{\cal O}_d LH}{\Lambda^{d-1}}, \\
{\cal O}_{d+3, B} & = & \frac{{\cal O}_d u^cd^cd^c}{\Lambda^{d}}, \nonumber \\ 
{\cal O}_{d+3, L1} & = &\frac{{\cal O}_d L L e^c}{\Lambda^{d}}, \nonumber \\ 
{\cal O}_{d+3, L2} & = &\frac{{\cal O}_d L Q d^c}{\Lambda^{d}}, \nonumber  \\
{\cal O}_{d+4} & = & \frac{{\cal O}_d LH LH}{\Lambda^{d+1}} \nonumber .
\label{higherdimW}
\end{eqnarray}
In supersymmetric theories it can be easier to satisfy observational
constraints on these operators, as rates for baryon- or lepton-number
changing processes can receive additional suppression from
superpartner mass scales.  

\subsection{Electroweak Sphalerons}

Baryon and lepton number are also broken nonperturbatively in the
SM by electroweak sphalerons.  To successfully transmit a
matter-antimatter symmetry from a hidden sector to the SM
via electroweak sphalerons requires a chiral mediator sector:
particles which carry both $SU(2)_L$ and the dark global symmetry,
such that the dark number symmetry becomes anomalous under $SU(2)_L$.
This enables electroweak sphalerons to reprocess a generated dark
asymmetry into a SM baryon (and lepton) asymmetry.  We
will present a model with a simple messenger sector in
section~\ref{sec:np} below.

In this scenario, the dark phase transition must again happen at
temperatures above the electroweak phase transition.  Moreover, the
messenger fields must now obtain all of their mass from electroweak
symmetry breaking, and therefore cannot be decoupled from the
electroweak scale.  Constraints on the messenger fields are therefore
more stringent than for perturbative mediation.  In particular,
precision electroweak constraints on additional heavy electroweak
matter as well as collider limits on direct production must be
avoided.

\section{Models of Dark Baryogenesis}
\label{sec:models}

We now present two explicit models of darkogenesis based on the
minimal supersymmetric dark hidden sector described in
section~\ref{sec:HS}: first, a low-scale model based on perturbative
mediation through the baryon-violating operator ${\cal O}_{d+3, B}$, and
second, a higher-scale model which uses electroweak sphalerons to
transfer the asymmetry.
The
common ingredient in both models is the supersymmetric version of the minimal dark hidden sector
described in section~\ref{sec:HS}, though the mass scale of the dark
symmetry-breaking phase transition and therefore of the dark Higgses
$H, H ^ c $ differs between the two models.  In the models presented
below, we incorporate a mechanism which generates the mass scale of
the dark Higgses dynamically, via a singlet which communicates to the
common origin of visible and hidden sector SUSY-breaking \cite{multi,Morrissey}.

Successful darkogenesis requires not only that a dark asymmetry be
generated and transferred, but also that the symmetric portion of the
DM abundance annihilate away.  We construct the Higgs
potential in the hidden sector to yield a spectrum which allows for
efficient annihilation of the symmetric DM abundance as well
as a first-order phase transition.

\subsection{Low Scale Dark Baryogenesis}
\label{sec:low}

In darkogenesis models where the dark phase transition occurs at
temperatures below the electroweak phase transition, the asymmetry
must be transferred directly to the baryons.  The lowest-dimension
neutral operator which can accomplish this is $\frac{1}{\Lambda^ p}
{\cal O}_d u^c d^c d^c$.  This operator must be in thermal equilibrium
at temperatures of order the dark phase transition, but
must leave equilibrium at temperatures $T_{dec}$ above where the DM becomes non-relativistic;
otherwise the transfer operator will wash out the dark asymmetry.  It
is easy to arrange this separation of scales $T_{dec} > m_{DM}$ in a
supersymmetric model, as any dark-visible interaction arising from the
superpotential term $W_{int} = \frac{1}{\Lambda^ d} {\cal O}_d u^c d^c
d^c$ must involve at least one squark.  This gives rise to a Boltzmann
suppression in the rate for the operator, which causes it to decouple
rapidly below the superpartner mass scale.  Thus while there is no
need for $B$ violation through rapid electroweak sphalerons in this
model, the typical scales for the dark phase transition are still
naturally related to the electroweak scale, through the mass scales
for SM superpartners.

To build a low-scale model, we connect the SM to the
supersymmetric minimal hidden sector of section~\ref{sec:HS} using the
operator
\begin{equation}
\label{eq:transfer}
W_{int} = \frac{1}{\Lambda^ 2} X ^ 2 u^c d^c d^c,
\end{equation}
where $X $ is the dark matter state.
We take this operator to be quadratic rather than linear in $X$ to
avoid inducing $X$ decay.  This interaction can be generated by
integrating out (for instance) a vector pair of color triplet
superfields $\zeta,\bar\zeta$ and a pair of singlet superfields $N,\bar
N$, with the renormalizable interactions
\begin{equation}
W = m_\zeta \zeta\bar\zeta + m_N N\bar N + d^ c d ^ c\zeta + \bar\zeta u ^ c N + \bar N X X .
\end{equation}
Again, $U(1)_D$-breaking mass terms for the $U(1)_D$-charged singlets
$N,\bar N $ must be forbidden. 
If one of the squarks is light, $m_{\widetilde q}\sim 200\gev$, the
decoupling temperature for this operator can be quite low: for DM masses
$m_{\widetilde X}
\sim m_{X}= 10\gev$ and taking $\Lambda\sim$ TeV, 
$\widetilde X \widetilde X \to q q \squark$ drops below the expansion
rate of the universe at $T\sim 50$ GeV.

The relation between the baryon number asymmetry $B $ and the dark number
asymmetry $D $ can be determined using the standard methods outlined in \cite{harveyturner}.
If the transfer operator freezes out after electroweak sphalerons have decoupled,
we find
\begin{equation}
\frac{B}{D}=\frac{23}{21},
\end{equation}
taking for concreteness one (Dirac) DM state and its superpartners in the
thermal plasma.  The DM mass is then determined to be
\begin{equation}
m_X =5 \frac{B}{D} m_p \approx 5 \gev .
\end{equation}
Out-of-equilibrium decays of the SM NLSP through the transfer 
operator could potentially alter this relation, but in our model do not, as we
will see below.
In the minimal weakly-coupled hidden sector the DM mass is
controlled by a technically natural Yukawa coupling and can be freely
adjusted.  However, as the masses of all light states in the hidden sector
are parametrically given by their couplings times the scale of dark symmetry breaking,
once the scale of the phase transition and the mass of the dark matter have been specified
the strength of the hidden sector interactions are no longer adjustable.

In order to show that the Sakharov criteria can be satisfied via our
superpotential, Eq.~(\ref{darksuper}), we must examine some of the
details of the phase transition in the hidden sector.  Rather than
setting the mass scale $\mu_D$ by hand, we generate it dynamically via
singlet mediation \cite{multi,Morrissey} which communicates SUSY
breaking from a GMSB messenger sector to the hidden sector.  This
mechanism also provides the means to radiatively break the dark
$SU(\chi)$ in the hidden sector.  We discuss this model as an example
of how one could successfully carry out dark baryogenesis in the
hidden sector given by Eq.~(\ref{darksuper}).  Other models could be
constructed.  We take the messenger scale to be sufficiently low that
the gravitino is lighter than the lightest state in the hidden sector.
We will see that the hidden Higgs potential thus generated can
naturally have a first-order phase transition.

To this end, we replace the dark $\mu $ term, $\mu_D H H^c$, with
the singlet terms
\begin{equation}
\label{eq:ws}
W_{dh} = \lambda S H H^c + \frac{\kappa}{3} S^3.
\end{equation}
We assume that $S$ obtains a weak scale SUSY breaking mass, and
furthermore that the soft mass squared for the scalar is positive.  It
is possible to achieve this SUSY breaking pattern via coupling to a SUSY-breaking
sector if the couplings are $R$-symmetric. 
No bare $B/\mu$ is generated, and $A$-terms in the potential $\lambda
A_\lambda S H H^c$ and $\kappa A_\kappa S^3$ only arise through
renormalization group flow.  Negative soft mass squareds are then
obtained for the dark Higgses $H,~H^c$ through one loop diagrams:
\begin{equation}
m_H^2 = m_{H^c}^2 \simeq -\frac{2 \lambda^2}{16 \pi^2} m_S^2 \ln\left( \frac{\Lambda}{m_{hid}}\right),
\end{equation}
where $m_{hid}$ is the mass scale of the hidden sector, and
$\Lambda $ is the scale where the singlet mass is generated.

Without the presence of an additional  dark quantum number, the two Higgs doublets of the hidden sector are indistinguishable.
In that case the general soft terms will then contain arbitrary mixings between
the doublets,
$
\mc{L}_{dh, gen} = m^2_{ij} \bar H_i H_j,
$
where $H_i, H_j = H, H ^ c $.
These more general Higgs quadratics complicate the minimization of
the potential.  To simplify the discussion we impose a global symmetry on the hidden Higgs potential so that it reduces to 
an (N)MSSM-like form.  There is, however, no problem in principle with allowing all terms mixing the Higgses.
Imposing the symmetry poses no cosmological problems, as it is
explicitly broken by the Yukawa couplings, preventing the formation of domain walls.
As in the NMSSM, with this potential only one component of each of dark Higgs (call it ``0'') will obtain a 
vev, and these vevs are located in opposite isospin components.  The 
potential for these degrees of freedom can be written
\begin{eqnarray}
V &=& \lambda^2 |S|^2(|H_0^c|^2+|H_0|^2)+\lambda^2|H_0^c H_0|^2+\frac{g_D^2}{8}(|H_0|^2-|H_0^c|^2)^2 + m_H^2 |H_0|^2 + m_{H_0^c}^2 |H^c|^2\nonumber \\ 
  & &  \phantom{spacer}
      + m_S^2 |S|^2 +\kappa |S| ^ 4 +\left(-\lambda\kappa H_0 ^ cH_0 \bar S ^ 2- \lambda A_\lambda S H^c_0 H_0 + \frac{\kappa}{3} S^3 +\rm{H.c.}\right).
\end{eqnarray}
The symmetry breaking pattern is
\begin{equation}
\langle H_0 \rangle^2 \simeq \langle H_0^c \rangle^2 \equiv \eta^2 \simeq -\frac{m_H^2}{\lambda^2}
\end{equation}
and
\begin{equation}
\langle S\rangle\equiv s = \frac{\lambda A_\lambda \eta^2}{m_S^2}.
\end{equation}
The minimum is stable provided $g_D^2 - 2\lambda^2 > 0$.

We now discuss the spectrum of the hidden sector in more detail to
ensure that there are no cosmological issues.  There are five physical
Higgses associated with $H,~H^c$ and two Higgses associated with $S$.
The real component of $S$ remains heavy, $m_{h_s}^2 \simeq m_S^2$.
There are three nearly degenerate Higgses (corresponding to one
``neutral'' Higgs and two ``charged'' Higgses) with masses $m_{h_1}^2
\simeq (g_D^2 - 2 \lambda^2) \eta^2$, and one lighter higgs with mass
$m_{h_2}^2 \simeq 2 \lambda^2 \eta^2$.  In the pseudo-scalar sector,
the theory has a global symmetry in the limit $\kappa \rightarrow 0$
or $A_{\lambda,\kappa} \rightarrow 0$, which is spontaneously broken
by $\eta$, so that there is a Goldstone boson.  The pseudoscalar
masses are the mostly singlet $m_{a_s}^2 \simeq m_S^2$ and the mostly
doublet $m_{a_h}^2 \simeq 6 \frac{s^2}{\eta^2}(-3 \lambda \kappa
\eta^2+\kappa A_\kappa s)$ (see for example \cite{Morrissey,Dobrescu}
for details).

\begin{figure}
  \begin{center}
    \setlength{\unitlength}{10 cm}
    \begin{picture}(0.7,0.4)
      \thicklines

      \multiput(-0.1,0.48)(0.02,0){6} 
      {\line(1,0){0.01}} 
      \multiput(0.07,0.48)(0.02,0){6}
      {\line(1,0){0.01}}  
      \put(-0.43,0.47){$\sim 100 \mbox{ GeV}$} %
      \put(0.3,0.48){$S,~a_s$}
      
       \multiput(0.07,0.4)(0.02,0){6} 
      {\line(1,0){0.01}} 
       \multiput(-0.1,0.4)(0.02,0){6}
      {\line(1,0){0.01}} 
      \put(-.1,0.39){\line(1,0){0.12}}
      \put(0.07,0.39){\line(1,0){0.12}}
      \put(0.3,0.4){$h_1,~h_\pm,~M_{1,2}^0,~M^\pm$}
      \put(-0.43,0.36){$\sim 10 \mbox{ GeV}$} %
      
      \multiput(0.07,0.33)(0.02,0){6} 
      {\line(1,0){0.01}} 
      \put(-.1,0.33){\line(1,0){0.12}}
      \put(0.3,0.33){$h_2,~M_{3,4}^0$}
      
      \multiput(-0.1,0.22)(0.02,0){6} 
      {\line(1,0){0.01}} 
      \put(-0.43,0.22){$\lsim 1 \mbox{ GeV}$} %
      \put(0.3,0.22){$a_h$}

      \put(0.85,0.34){\line(1,0){0.14}} 
      \put(0.71,0.34){$\gsim \mbox{GeV}$} %
      \put(1.0,0.34){$X$}

      \put(0.2,0.1){\line(1,0){0.3}} 
      \put(0,0.1){$<< \mbox{GeV}$} %
      \put(0.6,0.1){$\tilde{G}$}

    \end{picture}
  \end{center}
  \caption{The spectrum of the minimal dark sector.  Among the states
  carrying $U(1)_D$, only the lightest, $X$ (the DM particle), is shown.} \label{higgsspectrum}
\end{figure}
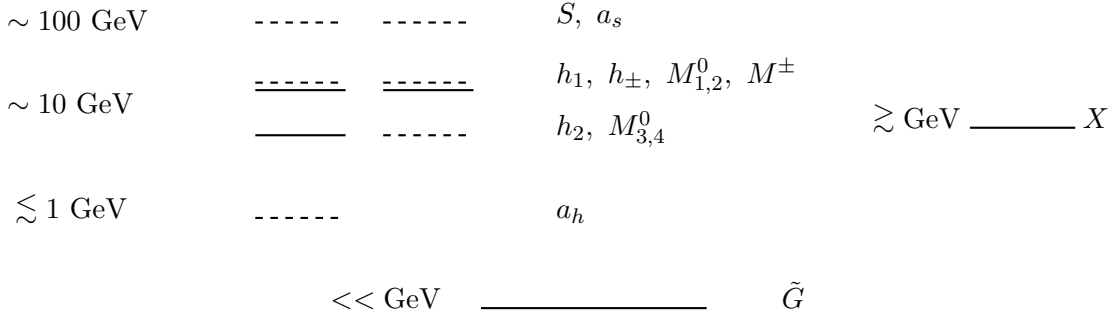

The neutralino mass matrix, in the limit $s\ll\eta$, is (in the $(\tilde{\lambda},
\tilde{H},\tilde{H}^c,\tilde{S}$) basis)
\begin{equation}
{\cal M}^f =\frac{1}{\sqrt{2}}\left( 
\begin{array}{cccc}
  0 & g_D \eta  &  -g_D  \eta & 0 \\
 g_D \eta & 0 & 0 & \sqrt{2} \lambda \eta \\
-  g_D  \eta & 0 & 0 & \sqrt{2}\lambda \eta \\
0 & \sqrt{2}\lambda \eta & \sqrt{2}\lambda \eta & 0
\end{array}
\right),
\end{equation}
giving two fermions with mass $M_{1,2}^0 = g_D \eta$ and two fermions
with mass $M_{3,4}^0 = \sqrt{2}\lambda \eta $.  These are nearly
degenerate with the Higgses $m_{h_1}$ and $m_{h_2}$.  In addition, the
charginos have masses $M^\pm = g_D \eta$.  A sample spectrum is shown
in Fig.~(\ref{higgsspectrum}).

The DM candidate is the lightest of the states carrying
$U(1)_D$, which we will henceforth denote $X $.  The mass
of the DM state is largely controlled by the Yukawa couplings
in the superpotential of Eq.~\ref{darksuper}, with mass splittings
within the chiral multiplets subdominant.  Since $H,~H^c$ receive
negative soft mass-squareds, within each matter multiplet the scalar
will be heavier than the fermion.  This implies that the DM
$X$ is fermionic, while its superpartner $\tilde{X}$ will decay to $X$
and the gravitino on a timescale
\barray
\tau (\widetilde X\to X \widetilde G ) & \simeq & \frac{16 \pi \langle F \rangle^2 m_X^3}{ \Delta m_X^8 } \nonumber \\
& \simeq & (3\times 10 ^ 4 \mbox{ s}) \left(\frac{\sqrt{F}}{100\tev}\right) ^ 4 \left(\frac{m_X}{10\gev}\right) ^ 3
              \left(\frac{10^{-3} (10\gev) ^ 2}{\Delta m_X^2}\right) ^ 4,
\earray
where $\Delta m^2_X\ll m^2_X$ is the mass-squared splitting between
the scalar and fermion induced by radiative corrections,
\barray
\Delta m_X^2 = -\frac{2 y^2}{16 \pi^2} m_H^2 \log \left(\frac{M_{mess}}{m_X}\right).
\earray
This decay should have a negligible effect on cosmology, however,
because both $X$ and $\widetilde{X}$ have efficient annihilations to the
light pseudoscalar $a_h$.

The light pseudoscalar provides an efficient annihilation mechanism
for removing the symmetric abundance of these particles.  Let $y $
denote the effective Yukawa coupling of the DM state to the
light pseudoscalar, $-{\cal L}_{int} \equiv i y a_h \bar X\gamma ^ 5 X$.
Then the annihilation cross-section for the process $\widetilde{X}^* \widetilde{X} \to a_h a_h $
is
\barray
\langle  \sigma v \rangle & \simeq & \frac{y^4}{64 \pi} \frac{1}{m_X^2} \nonumber \\
& \simeq & (2 \times 10^{-24} \mbox{ cm}^3/\mbox{s}) \left(\frac{y}{0.25}\right)^4 \left(\frac{\mbox{10 GeV}}{m_X}\right)^2 ,
\earray
which is sufficiently large to efficiently remove the symmetric 
component.

The axion itself will be stable in the absence of any interactions
with the SM.  Since it is massive, and its evolution thermal, its abundance 
could be problematic cosmologically.  On the other hand, if
it has a small coupling to the SM through a term $\Delta W = \zeta S
H_u H_d$, it can decay via the SM Higgs to photon pairs.  The lifetime
for this decay is as given in \cite{Gunion:1988mf, Morrissey}
\barray
\tau & \simeq & \frac{256 \pi^3}{\alpha^2 \lambda^2 \zeta^2} \frac{1}{F(m_{a_s},\tan\beta)^ 2}  {m_A^4}{\eta^2 m_{a_s}^3} \nonumber \\
& \simeq & (0.003 \mbox{s}) \left(\frac{10^{-3}}{\zeta}\right)^2 \left(\frac{0.1}{\lambda}\right)^2 
                   \left(\frac{ 225}{F(m_{a_s},\tan\beta)^2}\right) \left(\frac{m_A}{100 \mbox{ GeV}}\right)^4  \nonumber \\
    &  &   \:\:\:  \times      \left(\frac{40 \mbox{ GeV}}{\eta}\right)^2\left(\frac{m_{a_s}}{0.1 \mbox{ GeV}}\right)^3,
\earray
where $m_A$ is the MSSM pseudoscalar mass, $\tan\beta $ is the ratio
of MSSM Higgs vevs, and $F(m_{a_s},\tan\beta)$ is obtained by summing
over the contribution of SM fermions,
\begin{equation}
F(m_{a_s},\tan\beta) = \sum_i N_{ci} Q_i ^ 2 \left(-2 \frac{4 m_i ^ 2}{m_{a_s}} \right) f (4 m_i ^ 2/m_{a_s}) \left\{\begin{array}{cl} \cot\beta & \mbox{  up-type}\\
                          \tan\beta & \mbox{  down-type}      \end{array}\right. ,
\end{equation}
with
\begin{equation}
f (\tau) =\left\{\begin{array}{cc} (\sin ^{-1} \left(1/\sqrt{\tau})\right) ^ 2 & \tau\geq 1\\  
                                   -\frac{1}{4}\left(\ln \frac{1+\sqrt{1+\tau}}{1+\sqrt{1+\tau}}-i\pi\right)  ^ 2  & \tau< 1      \end{array}\right. .
\end{equation}
The coupling to the visible sector through $\zeta SH_uH_d$ can be efficient enough
to allow the axions to decay before they come to dominate the total energy density. We find that this is
sufficient for a 0.1 GeV axion for $\zeta  \gsim 10^{-3}$.

The small symmetric coupling through the Higgs portal also affects the
decay of the SM NLSP.  For $\zeta \gsim 10^{-3}$, if
the SM NLSP contains any $\widetilde W_3$ admixture, the decay
to a hidden sector Higgs and higgsino will dominate over both its  
decay to gravitino as well as potential baryon- and dark-number violating decays
through the transfer operator.  This ensures the symmetric abundance
of both $B$ and $D$ is not repopulated; meanwhile, the hidden sector higgsino can
decay rapidly to the axion and the gravitino.

In order to understand whether a nonzero dark number asymmetry is
generated during the phase transition we must consider the scalar
potential at finite temperature and verify that this spectrum is
consistent with a first-order phase transition.  A complete
calculation of the order of the phase transition would require a
careful numerical study; to obtain a basic estimate of the
requirements on the hidden sector couplings we perform a simplified
approximate analysis.  We work below the heavy singlet scale $m_s $,
and neglect terms which are higher order in the small $S$ vev.
Moreover, we make the simplifying assumption that the ratio of hidden
sector Higgs vevs does not substantially change as a function of
temperature, and consider only fluctuations in the linear combination
of fields responsible for gauge symmetry breaking, $\phi = \sqrt{ |H|
^ 2+|{H^c}|^2}$.  Neglecting the terms which involve the small $S$
vev, the zero-temperature potential for this degree of freedom is then
simply
\begin{equation}
\label{eq:v0limit}
V = m_H^2 \phi^2 + \frac{\lambda^2}{4}\phi^4,
\end{equation}
where $m_H^ 2 \equiv m_{H_0}^2 = m_{H_0^c} ^ 2 $.
The finite temperature thermal potential is given by
\begin{equation}
V (\phi, T) = V_0(\phi)+V_1 (\phi, T) + \ldots,
\end{equation}
where $V_1 (\phi, T) $ is the one-loop contribution.  As
$g_D>2\lambda$, the leading contribution to $V_1 (\phi, T) $ is from
the (transversely polarized) gauge bosons, which to leading order give
\barray
\label{eq:v1loop}
V_1 ^ {(\mathrm{gauge})} (\phi, T)&=&  \frac{g_g }{24}m ^ 2(\phi) T ^ 2 -\frac{g_g}{12\pi}(m ^ 2(\phi) ) ^ {3/2} T  +\ldots \\
\nonumber
         & = & \frac{1}{8}g_D ^ 2\phi ^ 2 T ^ 2 -\frac{1}{4\sqrt{2}\pi} g_D ^ 3\phi ^ 3 T+\ldots ,
\earray
where $g_{g}$ counts the number of (transverse) degrees of freedom.
Adding this one loop piece (\ref{eq:v1loop}) to the zero-temperature 
potential (\ref{eq:v0limit}) gives
\beq
\frac{\sqrt{2} \langle\phi (T_c)\rangle}{T_c} = \frac{g_D ^ 3}{2\pi\lambda^2}.
\eeq
The gauge bosons by themselves are then sufficient to drive the dark
phase transition to be sufficiently strongly first-order provided $\lambda $ is sufficiently
small, $ g_D^ 3/ 2\pi\gsim \lambda^2$.

Direct detection in this model is controlled by the small symmetric coupling 
through the Higgs portal.  In the limit of large $\tan \beta$, the scattering cross-section 
per nucleon is
\begin{eqnarray}
\sigma & \simeq & \frac{\mu_r^2}{\pi} N_n^2 y_h^2 \left(\frac{\lambda \zeta v_u \eta}{m_h^2}\right)^2
           \frac{1}{m_{h_2}^4} \nonumber \\
& \approx& 2 \times 10^{-43} \mbox{ cm}^2 y^2 \left(\frac{\mu_n}{m_p}\right)^2 \left(\frac{N_n}{0.1}\right)^2 
                    \left(\frac{\lambda}{0.1} \right)^ 2   \nonumber \\ &&\times 
    \left(\frac{\eta}{20 \mbox{ GeV}}\right)^2 \left(\frac{\zeta}{10 ^{-3}} \right)^ 2 
                  \left(\frac{115\gev}{ m_h} \right)^ 4 \left(\frac{10\gev}{ m_{h_2}} \right)^ 4,
\end{eqnarray}
where $\mu_r$ is the DM-nucleon reduced mass, $y_h \sim m_D/\eta$ is the effective Yukawa coupling of the
dark matter to the light dark Higgs eigenstate, and $N_n$ is the coupling of the
MSSM Higgs to the nucleons.  This is in reach of direct detection experiments.

DM can additionally scatter off visible matter via the
baryon-number violating transfer operator $W_{int} =\frac{1}{\Lambda^ 2} X^2 u^c d^c d^c$.  
In the present model, since the splitting between
the DM state and its superpartner is smaller than the
proton-pion mass splitting, the dominant baryon-violating scattering
process is
\begin{equation}
\label{eq:pdmscatter}
p^ + X\to  \widetilde {\bar X}\pi ^ + .
\end{equation}
This process must proceed through a loop of SM superpartners.  The
rate for this process is on the order of $10 ^{-30}/$year for
TeV-scale superpartners.  While the present bounds on the proton
lifetime are 8.2$\times 10^{33}$ years (for $p ^ +\to e ^ +\pi ^ 0$) and
6.6$\times 10^{33}$ years (for $p ^ +\to \mu ^ +\pi ^ 0$) \cite{:2009gd},
these limits are not applicable to the process of
Eq.~(\ref{eq:pdmscatter}), where the pion is the only visible particle
in the final state.

\subsection{Mirror Messengers and High Scale Dark Baryogenesis}
\label{sec:np}

If the dark phase transition occurs above the electroweak phase 
transition, a generated dark asymmetry can be communicated to the SM
via electroweak sphalerons, instead of through
higher dimension operators.  This mechanism requires the introduction of
messenger fields which carry both $SU(2)_L$ and $U(1)_D$ quantum
numbers, such that $U(1)_D$ becomes anomalous under $SU(2)_L$.
We will call these chiral messengers leptodarks, as in our model 
they will have lepton-like SM charges, in addition to carrying
dark number.

In order to avoid fractionally charged states after electroweak
symmetry breaking, we must either ensure any fractionally charged 
leptodarks are bound into integrally charged composites, as in 
\cite{quirk}, or assign hypercharge to the leptodarks in such a 
way that the resulting states after electroweak symmetry breaking 
are integrally charged.  This ensures that the lightest messenger 
can decay.   

A minimal chiral messenger sector is shown in Table~(\ref{table:mess}).
The messenger sector carries
vector-like SM quantum numbers, ensuring anomaly
cancellation, and chiral $U(1)_D$ quantum numbers.  The hypercharge 
assignments are necessary to ensure that all states after EWSB have
integral charge. The superpotential in this messenger sector takes on the form
\begin{equation}
W_M = y_{eM+} L_M ^ + H_u {e_M^c}^+ + y_{eM-} L_M ^- H_d {e_M^c}^ - +  y_{XM+} L_M ^ + H_d \bar{X}_M^i + y_{XM-} L_M ^- H_u \bar{X}_M^i.
\end{equation}
The messengers must also carry the same global quantum number as the
hidden sector $U(1)_D$.

The messenger fields contribute to precision electroweak observables.
In the limit where the mass splitting between charged and neutral
leptodarks goes to zero, the messenger contribution to precision
electroweak observables is
\begin{equation}
\Delta S = \frac{1}{3\pi}\simeq 0.11,\phantom{spacer} \Delta T\simeq \Delta U \simeq 0,
\end{equation}
which is compatible at 95\% CL with observations \cite{lepewwg,
Kribs:2007nz}.  Agreement with data can be further improved by
adjusting the mass splittings between the components of the leptodark multiplets.  The
most stringent collider constraints on the messengers are the LEP mass
limits, $m_{L^\pm}> 100.8 \gev$ for charged leptodarks and $m_{N}
>45.0\gev $ for neutral leptodarks \cite{Kribs:2007nz, pdg}.

The details of the dark Higgs potential can be taken to be the same
as the singlet-mediated example for the low-scale model, with the
overall mass scale translated to values above the weak scale
to trigger the early phase transition.  This can occur if the soft
singlet mass is now significantly above the weak scale, a viable
option depending on the couplings of the singlet to the source of SUSY-breaking.
Raising the intrinsic hidden sector scale so far above the dark matter mass scale
means, in our minimal sector, that the interactions of the light hidden
sector degrees of freedom become weak, and generically additional structure 
is required to remove the thermal relic abundance.
Communication between the dark and messenger sectors as the electroweak sphalerons freeze out 
can be ensured either by allowing a significant mixing between the messenger and dark singlets, or 
by introducing an additional explicit coupling 
$
W_{int} =\xi Z \bar X_D X_M
$
involving a singlet superfield $Z$, and where we have added a subscript to the dark singlets 
$\bar X_D$ for clarity.  In this latter case, the $U(1)_D$ quantum numbers
for the messenger fields are reversed from those chosen in Table~II.

The DM is again fermionic, and consists of a
dark singlet $\bar X_D$ paired with one component of a dark doublet 
$L_D$ with possible admixtures from the neutral component of a 
messenger doublet $L_M$ controlled by the mixing angle $\sin \chi\approx \chi$. 
To avoid constraints from the invisible $Z^ 0$ width, 
the messenger  doublet component in the DM state 
should be small, $\chi^ 4 \lsim 10 ^{-3}$.

This model, having a high scale phase transition, suffers from a greater weakness than the low 
scale model because there are naturally no additional light states present for the dark state 
to annihilate efficiently to.  One can remedy this by introducing additional light degrees of 
freedom in the hidden sector.  For example, one can introduce
an additional pair of matter multiplets in the hidden sector, $Y$ and $\bar Y$, which 
are $SU(2)_D$ singlets, carry
dark number $+1$ and $-1$ respectively, and interact with the DM through
\beq
W_Y = m_Y \bar Y Y+ \xi_Y Z \bar X_D Y .
\eeq
This interaction allows the dark matter to annihilate to light $Z$ fermions with a 
cross-section
\barray
\label{eq:symmannih}
\langle \sigma v \rangle (X \bar X \to \widetilde z\widetilde z) &= &\frac{\xi_Y ^ 4}{32\pi m_X ^ 2}\left(\frac{m_X}{m_Y}\right) ^ 4 \\
& = & 2\times 10 ^{-24} \mbox{cm} ^ 3/\mbox{s} \:\: \xi_Y ^ 4\left(\frac{m_X}{\mathrm{GeV}}\right)^ 2 \left(\frac{15\gev}{m_Y}\right) ^ 4.
\nonumber
\earray
The scalars $\widetilde X$ can decay to $X \widetilde G$ before the $X$ annihilation process freezes out.
The $Z$ scalar similarly decays to the massless fermionic $\widetilde z$, which is
stable.  As long as the $\widetilde z$ fermions are sufficiently light, this causes no 
problem with constraints on additional degrees of freedom from BBN, since the hidden 
sector is much cooler than the SM, 
having decoupled from the visible sector before the QCD phase transition.

With the introduction of the messenger $SU(2)_L $ doublets, the 
electroweak sphalerons now violate the global $U(1)$ number $B + L
+ \frac{N_D}{N_g} D$, where $N_g=3$ is the number of SM generations
and $N_D=2$ is the number of messenger electroweak doublets.  This
reprocesses the DM number asymmetry generated from the
dark phase transition into SM $B$ and $L$.  

The precise relation between the baryon asymmetry and the dark asymmetry
depends on a number of factors, such as how rapidly the sphalerons
decouple during the electroweak phase transition, whether the top
quark is integrated out of the theory when the sphalerons decouple,
and how many dark and messenger fields have their mass below the EWPT.  For
concreteness, we take the top quark as well as all of the messengers to be
heavier than the sphaleron decoupling temperature, and we take only the DM 
state and its superpartner integrated into the theory below the
electroweak phase transition.  All other hidden sector states we integrate out. In
this case, assuming the electroweak phase transition is second-order,
we find the dark asymmetry is related to the baryon asymmetry
as
\begin{equation}
\frac{B}{D} = \frac{33}{127},
\end{equation}
predicting a DM mass of approximately 1 GeV.

\begin{table}
\begin{center}
\begin{tabular}{cccc}
\hline \hline
& $SU (2)_L$ & $U(1)_Y $ & $U(1)_{D} $\\
\hline

$L_{M}^\pm$      & 2 & $\pm\frac{1}{2} $ & $1$\\
${e_{M}^c}^\pm$     & 1 & $\mp 1$               & $-1$ \\
$\bar{X}_{M}^i$     & 1 & $0$               & $-1$ \\
\hline \hline
\end{tabular}
\caption{A minimal dark messenger
sector\label{table:messengercontent}.  Anomaly cancellation is
achieved via mirror fermions with the same $U(1)_{D}$ charge, but
opposite hypercharge.  There are two sterile states,
$\bar{X}^{1,2}_M$.
\label{table:mess}
}
\end{center}

\end{table}

With the introduction of the messenger electroweak doublets, direct detection can now
proceed via the doublet fraction of the DM state as well as through mixing between the
hidden and visible Higgses.  The DM-neutron cross-section 
from the doublet component of the DM state scattering through the $Z^ 0$ is, 
\begin{eqnarray}
\sigma &=& \frac{G_F^2 \chi^4 \mu_r^2}{512 \pi} \nonumber \\ 
&\approx&5 \times 10^{-42} \mbox{ cm}^2 \left(\frac{\mu_r}{m_p}\right)^2 \left(\frac{\chi}{0.1}\right)^4,
\end{eqnarray}
where $\chi$ is again the effective coupling of the DM to the $Z^ 0$ via mixing with the messenger doublets.
This is large enough to be constrained by monojet searches at the Tevatron \cite{TeV}.

\section{Summary and General Comments about Darkogenesis}
\label{sec:conclusion}

Our purpose in this paper was to outline the general requirements of a
dark sector which can accomplish dark baryogenesis via a first
order phase transition in the hidden sector.  
We constructed a minimal weakly coupled 
dark sector which generates a matter-antimatter asymmetry, and discussed messenger
sectors for transferring the asymmetry to the SM.  We focused on a scenario
where dark number violation and out-of-equilibrium dynamics in the
hidden sector can be achieved via a dark non-Abelian gauge group with
a first-order-phase transition, under which dark number is anomalous.
The chiral anomaly of the dark number current provides $C$ violation, and $C$ and $CP $ 
are both violated through chiral couplings of dark
states to the dark Higgses.  In these models the DM mass lies
in a low mass window, between approximately 1 and 15 GeV, and the low 
symmetry-breaking scale for the dark sector can be generated dynamically via 
the mediation of SUSY breaking.  We
constructed an explicit dark Higgs sector which satisfies the
requirements for a strong first-order phase transition, and in addition
provides a mechanism for successful removal of all non-asymmetric relics in the
dark sector.  The minimal weakly-coupled hidden sector which is the basis of our models has the
special feature that the interaction strengths of all light states are
determined by the ratio of their mass to the symmetry-breaking scale and the hidden sector.  
As the dark matter must be sufficiently strongly interacting to remove its thermal relic abundance,
this structure requires low scales for the dark phase transition or 
additional structure in the dark sector.
We showed that when this phase transition occurs below the electroweak
phase transition (so that baryon number violation through electroweak
sphalerons is off), that the dark asymmetry can still be transferred
efficiently to the baryons via a higher-dimension operator.
Alternatively, the dark phase transition can occur well before the
electroweak phase transition.  In this case, a messenger sector which carries
both $SU(2)_L$ and the dark number can render dark number anomalous under 
$SU(2)_L$, thereby transferring the dark asymmetry into baryons.

Direct detection cross sections in these models depend on small symmetric
connections between the hidden sector and the visible sector, which have no
intrinsic connection to the darkogenesis mechanism or
the relation between the DM and baryon number.  
The gravitational wave signal from the first-order phase transition
could provide an orthogonal probe of our darkogenesis scenario, 
and for transition temperatures in the interesting range from below 100 GeV to 100 TeV is 
potentially within reach at upcoming gravitational wave 
observatories \cite{gravity}.

Many other interesting and viable scenarios of Darkogenesis remain to be
developed.  For example, $CP$ violation
could be introduced through the coupling of the dark states to the
visible sector, rather than of the non-Abelian
hidden sector to itself.  With the great freedom offered by nontrivial dark sectors, many further 
 novel avenues remain to explore.

\section*{Acknowledgements}
We thank  R. Gran for comments about Super-Kamionkande constraints, and M. Buckley, 
H. Davoudiasl, R. Harnik, G. Kribs, D. Morrissey, L. Randall, 
J. Thaler, W. Skiba and T. Volansky for useful comments
and discussions.  We thank the Aspen Center for Physics (JS and KMZ) and SLAC (JS) 
for hospitality during the final stages of this work.  JS was supported in part by DOE grant DE-FG02-92ER40704.

\end{document}